\documentclass[pra,twocolumn,showpacs]{revtex4-1}

\usepackage{amsmath,amssymb,graphicx}
\usepackage[utf8]{inputenc}
\usepackage[T1]{fontenc}

\begin{document}

\title{High quality photonic entanglement out of a stand-alone silicon chip}

\author{Dorian Oser$^1$, S\'ebastien Tanzilli$^2$, Florent Mazeas$^2$, Carlos Alonso Ramos$^1$, Xavier Le Roux, Gr\'egory Sauder$^2$, Xin Hua$^2$, Olivier Alibart$^2$, Laurent Vivien$^1$, \'Eric Cassan$^1$, and Laurent Labont\'e$^{2,}$}
\email{laurent.labonte@univ-cotedazur.fr}
\affiliation{$^1$Centre de Nanosciences et de Nanotechnologies, CNRS, Universit\'{e} Paris-Sud, Universit\'{e} Paris-Saclay, Palaiseau, France\\
$^2$Universit\'{e} C\^ote d'Azur, CNRS, Institut de Physique de Nice (INPHYNI), Parc Valrose, 06108 Nice Cedex 2, France
}

\begin{abstract}

The fruitful association of quantum and integrated photonics holds the promise to produce, manipulate, and detect quantum states of light using compact and scalable systems. Integrating all the building-blocks necessary to produce high-quality photonic entanglement in the telecom wavelength range out of a single chip remains a major challenge, mainly due to the limited performance of on-chip light rejection filters.  We report a stand-alone, telecom-compliant, device that integrates, on a single substrate, a nonlinear photon-pair generator and a passive pump rejection filter. Using standard channel-grid fiber demultiplexers, we demonstrate the first entanglement qualification of such a integrated circuit, showing the highest raw quantum interference visibility for time-energy entangled photons over two telecom-wavelength bands. Genuinely pure maximally entangled states can therefore be generated thanks to the high-level of noise suppression obtained with the pump filter. These results will certainly further promote the development of more advanced and scalable photonic-integrated quantum systems compliant with telecommunication standards.
\end{abstract}

\flushbottom
\maketitle

%\thispagestyle{empty}

% \section*{Introduction}
Quantum information science (QIS) exploits the fundamental properties of quantum physics to code and manipulate quantum states. QIS is regarded as the most promising pathway towards disruptive technologies, envisioning major improvements in processing capabilities and communication security \cite{bennett_quantum_2000, acin_quantum_2018}. However, practical implementations, such as quantum key distribution systems or quantum processors, require a large amount of compatible building-blocks \cite{knill, diamanti,qiang_large-scale_2018, samara_high-rate_2019}. Integrated photonics provides efficient and reliable solutions for realizing advanced quantum communication systems based on both linear and nonlinear elements \cite{grassani_micrometer-scale_2015,chen_wavelength-tunable_2016,aktas,0034-4885-80-7-076001, Wangeaar7053, PhysRevLett.113.103601, vergyris2016chip}. Still, all these realizations face a crucial limitation as soon as on-chip suppression of photonic noise is concerned due to the substantially higher pump intensity compared to that of the photon-pairs. Most of the time, this operation is externalized, using fiber or bulk optical components, and hinders the benefit of both the compactness and stability of the whole system \cite{Faruque:18}.

CMOS-compatible technologies hold the promise of bringing quantum photonics one step further with optical circuits showing higher integration levels \cite{silverstone2015qubit}. A few experiments based on this technology have been carried out to address the pump rejection challenge using on-chip solutions \cite{popovic_multistage_2006,luo_silicon_2012,dong_ghz-bandwidth_2010, piekarek_high-extinction_2017, harris_integrated_2014, ong_ultra-high-contrast_2013,elshaari_-chip_2017}. On-chip pump rejection has been demonstrated based on a semiconductor quantum dot integrated in a CMOS photonic circuit, but emitting out of the telecom range. The other strategies suffer from two main limitations: i) the continuous monitoring of the filter response to maintain proper performance\cite{popovic_multistage_2006,luo_silicon_2012,dong_ghz-bandwidth_2010}, and ii) prohibitive additional interconnection losses between components \cite{piekarek_high-extinction_2017, harris_integrated_2014} (up to 9 dB\cite{ong_ultra-high-contrast_2013}). Moreover, all these solutions have reported temporal correlation measurements for revealing the degree of indistinguishability between the paired photons \cite{piekarek_high-extinction_2017,harris_integrated_2014}. Qualifying quantum correlations in a stricter way, \textit{i.e.} by means of a Bell-type entanglement witness\cite{franson1989bell}, is essential for a large variety of quantum applications such as secret key distribution \cite{PhysRevLett.67.661}, superdense coding \cite{PhysRevLett.76.4656}, teleportation \cite{PhysRevLett.92.047904}, sensing \cite{kaiser_quantum_2018}, and computing \cite{qiang_large-scale_2018}. Such an entanglement characterization still remains unanswered whatever the integration platform (silicon, III-V semiconductors, lithium niobate) exploited for stand-alone devices embedding a pump-rejection solution. Yet, the degree of entanglement could be reduced by excess background noise or Raman photons induced inside the on-chip pump filter itself. 

In this work, we demonstrate and qualify a wavelength multiplexed entangled photon-pair source fully compliant with fiber telecom networks and semiconductor technology. Based on CMOS-compatible silicon photonics platform, our design includes a photon-pair generator and a pump rejection filter, paving the way for compact, low-cost, and telecom compliant quantum solutions. The rejection filter strategy has already shown in a previous work a pump suppression level up to 80 dB \cite{oser2018coherency}. Moreover, we proceed to the qualification of energy-time entanglement carried by the photon-pairs using a standard Franson-type interferometer \cite{franson}. We show two-photon interference fringes with a raw visibility exceeding 92\% over eleven complementary channels pairs spanning from both S and C telecom bands \cite{telecom} along with a coincidence-to-accidental ratio reaching 60. By complementary, we understand frequency correlated channels apart from the pump channel. The next step will consist in integrating wavelength demulitiplexing components directly on chip so as to obtain a fully functionalized and scalable entanglement supplier \cite{Matsuda:14}.

\section*{Results}
\paragraph*{Photonic device \& classical measurements}

\begin{figure*}
	\centering
	\includegraphics[width=\linewidth]{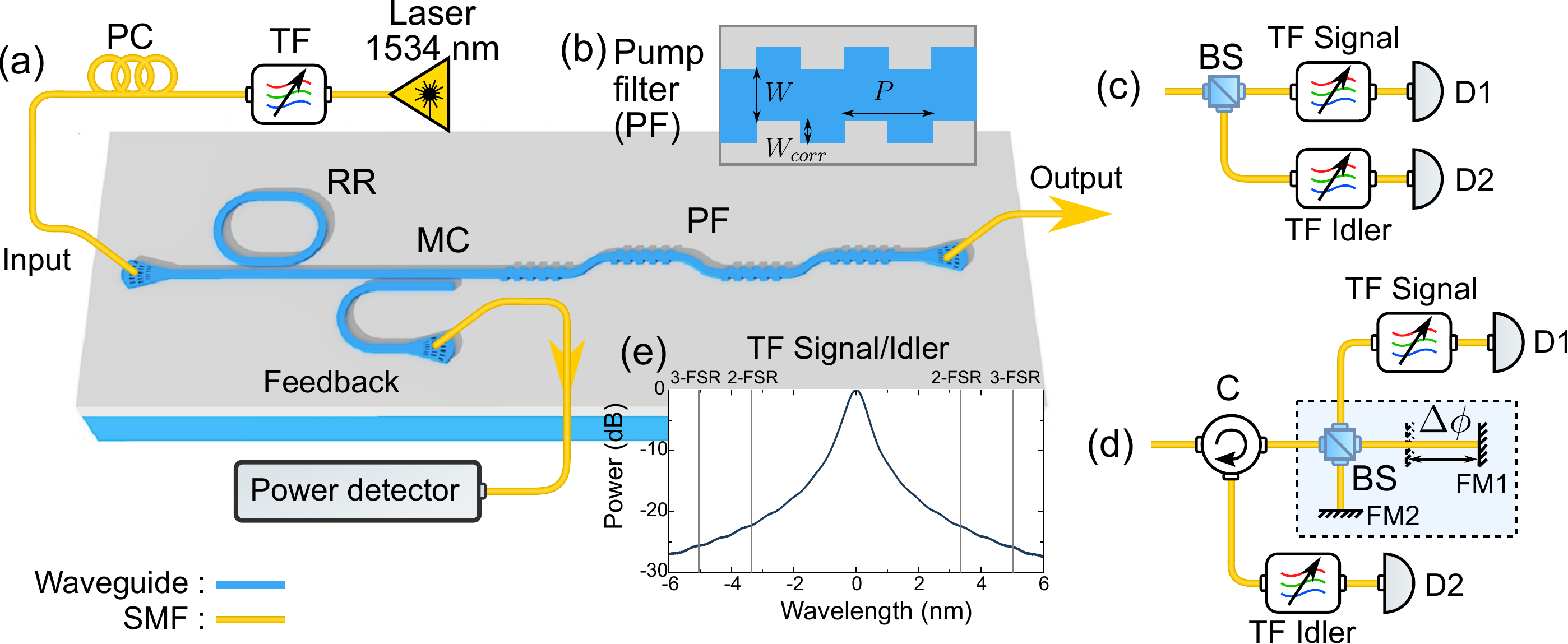}
	\caption{Schematics of the experimental setup. (a) Input laser with associated filter (TF) rejecting the amplified spontaneous emission and a polarization controller (PC). Laser light is injected in the ring resonator (RR) through the grating coupler, and then propagates to a modal coupler (MC) and through the integrated pump filter (PF). (b) Schematic top-view of one of the cascaded Bragg filter (BF) composing the PF. The output of the chip is connected to either, a coincidence setup (c) with a beam splitter (BS) and bandpass filters to demultiplex signal and idler photons (TF signal/idler), or to the entanglement qualification setup (d) in a folded-Franson configuration. (e) Spectrum of the signal and idler filters, which exhibit $22 \;\mathrm{dB}$ and $25 \;\mathrm{dB}$ rejection for the 2-FSR shift and 3-FSR shift configurations, respectively.}
	\label{fig:schema}
\end{figure*}

\begin{figure*}
	\centering
	\includegraphics[width=0.70\linewidth]{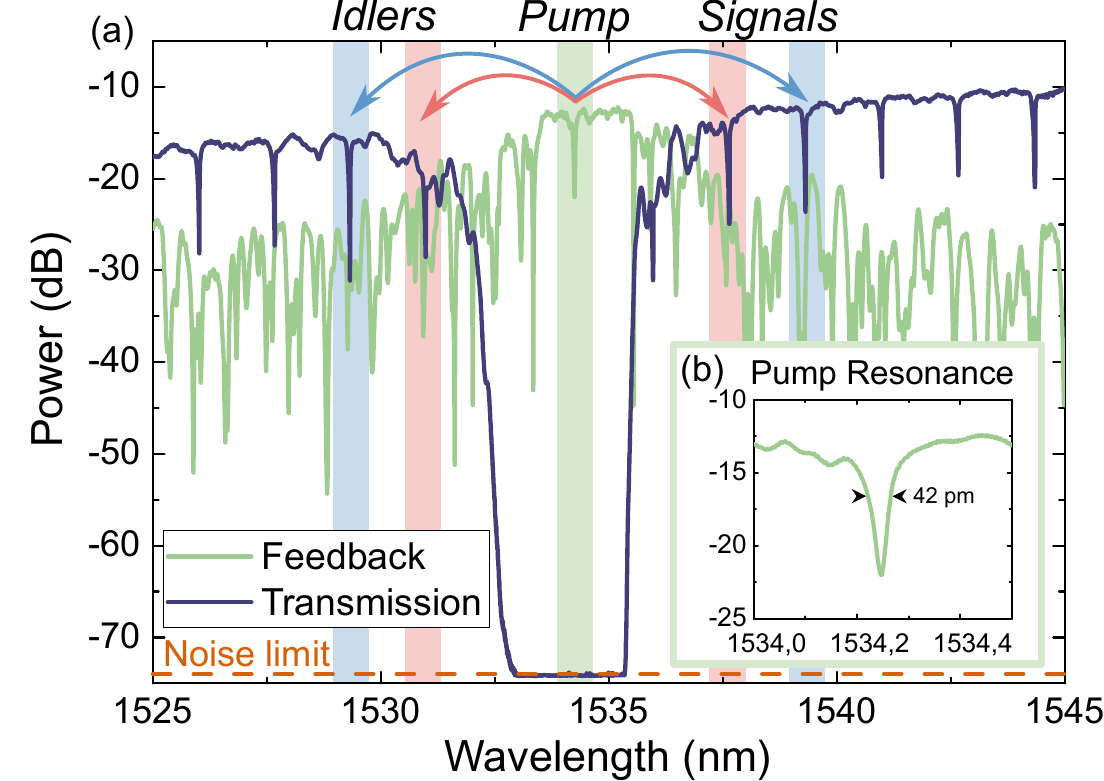}  % Small prints
	\caption{Transmission spectrum of the circuit. (a) Scan of the output port of the chip as well as the feedback port which represents the portion of light reflected by the first section of the filter. The filter exhibits a rejection of $60 \;\mathrm{dB}$ (limited by the noise floor of the detector). (b) Zoom over the pumped resonance of the ring. As can be seen, it corresponds to the center of the pump filter and its 3dB-bandwidth is $42 \pm 0.5  \;\mathrm{pm}$. Signal, idler, and pumped resonances used in the experiments are highlighted, in red the 2-FSR shift, blue the 3-FSR shift, green being the pumped resonance.}
	\label{fig:spect}
\end{figure*}

As sketched in Fig. \ref{fig:schema}a, our silicon chip integrates several building-blocks towards realizing a stand-alone quantum photonic device: a ring resonator (RR), a pump filter (PF) based on multi-mode cascaded Bragg filter, and modal couplers, as well as in and out grating couplers. The quality factor and the free spectral range (FSR) of the RR are equal to $Q = 3 \cdot 10^4$ and $200 \;\mathrm{GHz}$ around $1535 \;\mathrm{nm}$, respectively. As shown in Fig. \ref{fig:spect} (a), the transmission profile of our RR features by design a frequency-comb structure matching that of the two first telecom channel pairs (see Methods for the design and the fabrication). This makes it possible to use off-the-shelves telecom components for demultiplexing and routing the information out of the chip. Through spontaneous four wave mixing (SFWM), photons pairs are created according to the conservation of both the energy and the momentum. The photon pairs are produced in a symmetrical way with respect to the pump wavelength (Fig. 2 (a) and Fig. 2 (b)). In the following, long-wavelength photons are referred to as signal photons, whereas short-wavelength ones to as idler photons. Concerning the PF, the grating period of the Bragg filters has been chosen to reflect light around $1535 \;\mathrm{nm}$ into the second order mode (see Methods). By adding a single-mode waveguide between the sections of the PF, the second mode is radiated into the substrate, breaking the coherence that would otherwise been established. This strategy allows implement high-rejection filters being all-passive thanks to the cascading of modal-engineered Bragg gratings with relaxed fabrication requirements \cite{oser2018coherency}.

A modal coupler is added between the ring and the first Bragg section to recover part of the reflected light through the feedback port \cite{wang2015broadband}. This makes the alignment of the pump laser to the desired resonance easier even with thermal drift. Note that no cladding is added to improve the natural transverse magnetic (TM) polarization filtering of the waveguide. Finally, subwavelength fiber-chip grating couplers are employed to inject and extract transverse electric (TE) polarized light using standard single-mode optical fibers \cite{benedikovic2015subwavelength}. %These grating couplers are optimized to reduce Fabry-Perot ripples in the collected signal for an easier analysis of the transmission spectrum.

Prior to quantum qualification, we characterize the circuit in the classical regime by measuring its transmission spectrum using a tunable laser associated with a data acquisition system (Yenista Tunics and CT400) (Fig. \ref{fig:spect}). We use a polarization rotator to inject TE-polarized light into the grating. 
%This provides information on possible wavelength misalignment between the PF bandwidth and the frequency comb generated by the ring resonator. 
The pump rejection filter has a 3-dB bandwidth of $5.5 \;\mathrm{nm}$. In our case the usable bandwidth is only $3 \;\mathrm{nm}$, as only the deepest part of the filter can be exploited for the rejection of the pump. 
%Note that a metallic heater, deposited on top of the resonator, would have permitted fine alignment of the generated frequency response of the ring and filter.
As shown in Fig. \ref{fig:spect}, the measured rejection of the filter is higher than $60 \;\mathrm{dB}$ (dark-blue curve), which corresponds to the noise floor (red dashed line) of our detector. This measured rejection rate is consistent with state-of-the-art realizations for single-chip PF \cite{popovic_multistage_2006,luo_silicon_2012,dong_ghz-bandwidth_2010}. To further investigate the performances of the PF, we need to qualify the entangled states generated on an advanced integrated circuit including this PF.

\paragraph*{Time correlation measurements}

\begin{figure*}

\includegraphics[width=\linewidth]{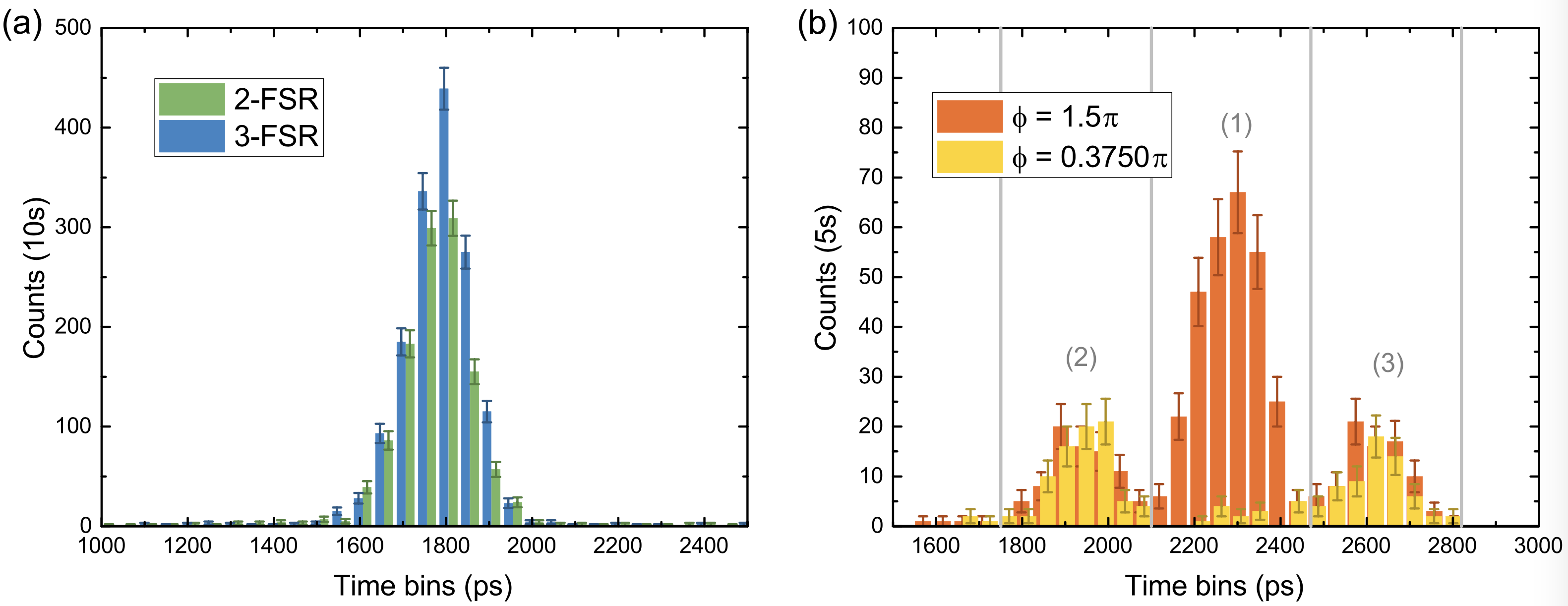}
%	\centering
%	\begin{minipage}{.5\textwidth}
%	  \centering
%	  \includegraphics[width=\linewidth]{FigHistoCoins.eps}
%	\end{minipage}%
%	\begin{minipage}{.5\textwidth}
%	  \centering
%	  \includegraphics[width=\linewidth]{FigHistoInterf.eps}
%	\end{minipage}
	\caption[width=\linewidth]{Recorded coincidence histograms. (a) The histograms represent the recorded coincidences for both 2-FSR shift and 3-FSR shift resonances. The overall integration time is 10 seconds and the measurements are done with SSPD detectors (Fig. \ref{fig:schema}b). The 2-FSR rate is of about $116$ counts/s with an SNR of 67.5, while the 3-FSR coincidence rate is of about $151$ counts/s with a SNR of 100. The noise level within the peak is between $1.5$ counts/s and $1.7$ counts/s. (b) The histograms represent the coincidences at the output of the interferometric setup sketched in  Fig.\ref{fig:schema}c for the 2-FSR shift resonances. They show a maximum and a minimum of interference (region (1)). Here, the overall integration time is of 5 seconds. The error bars for all points come from Poissonian statistics associated with the pairs (e.g. $\sqrt[]{N}$, $N$ being the number of coincidences). Note that similar histogram are obtained for the 3-FSR shift case.} 
	\label{fig:coins}
\end{figure*}

We proceed to time correlation measurements as a prerequisite to two-photon interference for qualifying the amount of entanglement carried by the photon pairs generated and filtered on chip. To reveal time correlation (Fig. \ref{fig:schema}c), signal and idler photons are separated and sent to different detectors that are connected to a time-to-digital converter (TDC). The statistic in the arrival times and related delays is then recorded and reconstructed as a coincidence histogram (Fig. \ref{fig:coins}a). To this end, the paired photons are demultiplexed at the chip's output with off-the-shelf filters of only 22 dB cross-talk (as shown in Fig. \ref{fig:schema}e).  The pump is set to $\lambda_p = 1534.2 \;\mathrm{nm}$ (C band) with an input power of $2.8 \;\mathrm{mW}$ after the polarization controller (Fig. \ref{fig:schema}a). As a first step, we study the closest resonances from the pump, \textit{i.e} paired channels distant by two and three FSR, as they may suffer preferably from the pump photonic noise. The related wavelength for the 2- and 3-FSR shift are highlighted in Fig. \ref{fig:spect} and detailed in the Methods. Examples of typical coincidence histograms are shown in Fig. \ref{fig:coins}a, where distinct coincidence peaks emerge over a small background of accidental counts. This stands as a clear signature of the simultaneous emission of the photon pairs. The width of the coincidence peak is given by the convolution of the coherence time of the photons, $\sigma_{coherence} \sim 110\;\mathrm{ps}$, of the detectors' timing jitters, $\sigma_{jitter} \sim 100\;\mathrm{ps}$, and of the time resolution of the TDC, $\sigma_{resolution} \sim 1\;\mathrm{ps}$. The full width at half maximum of the coincidence peak is about $160 \;\mathrm{ps}$, which is consistent with  $\sqrt{\sigma_{coherence}^2 + \sigma_{jitter}^2 + \sigma_{resolution}^2} \sim 150\;\mathrm{ps}$. Both the 2-FSR shift and 3-FSR shift resonances exhibit similar results.

We now consider the internal brightness of the ring. The overall losses of the setup, including the input/output and propagation losses of the chip, are outlined in Table \ref{tab:loss} of the Methods section. The single count rates in the coincidence experiment are of $4 \cdot 10^4$ signal counts per seconds and $3 \cdot 10^4$ idler counts per seconds, with $17 \;\mathrm{dB}$ and $18.5 \;\mathrm{dB}$ of losses, respectively. The overall coincidence peak spreads over several time bins and shows an average of 120 coincidences per second over a time window of $400 \;\mathrm{ps}$ (Fig. \ref{fig:coins}a). With all those figures, we can infer an internal generated photon-pair rate of $(2.1 \pm 0.2) \cdot 10^6 \;\mathrm{pairs/s}$. As the ring shows resonances of $42 \;\mathrm{pm}$ width (Fig. \ref{fig:spect}b) and the power in the ring is estimated to be $0.9 \;\mathrm{mW}$, we estimate an internal brightness of $\sim 500\;\mathrm{pairs/s/mW^{2}/MHz}$. Due to the non-deterministic wavelength separation induced by the beam splitter (Fig. \ref{fig:schema}a) and the spectral filtering ensured by the bandpass filters in each arm, the rate at the output of the chip is 4 times higher, \textit{i.e} about 480 pairs generated per second for each channel pair. Let us stress that this coincidence rate, stands among the best values reported for photonic devices embedding several key components \cite{harris_integrated_2014, azzini_ultra-low_2012, jiang_silicon-chip_2015, Wakabayashi:15, grassani_micrometer-scale_2015, suo_generation_2015,santagati_silicon_2017}. In comparison, similar realizations suffer from low coincidence rates due to prohibitive losses, precluding any further analysis of entanglement \cite{harris_integrated_2014,piekarek_high-extinction_2017}. Note that the other interesting feature reported in Table \ref{tab:loss} is the 2\;dB-loss for the pump filter which is almost only due to propagation. This low value associated with a high rejection level and a narrow bandwidth makes our pump filtering strategy a promising candidate for next generation quantum photonic circuits.

 Before addressing entanglement analysis, a relevant figure of merit associated with time correlation measurements consists in evaluating the signal-to-noise ratio (SNR) between the coincidence peak and the background noise.  In the measurements presented in Fig. \ref{fig:coins}a, the SNR is greater than 60 for both histograms (2-FSR shift and 3-FSR shift). Accidental counts mainly come from successive pairs events, when one of the two photons has lost its paired companion. This SNR can be improved by using a lower pump power at the price of reduced coincidence counts and of longer integration times\cite{Rogers}. There, our strategy is slightly different and promotes pragmatic realizations of QIS experiments by emphasizing high-coincidence counts while keeping a moderate but safe SNR~\cite{aktas}.

\paragraph*{Energy-time entanglement analysis}

\begin{figure*}
\includegraphics[width=\linewidth]{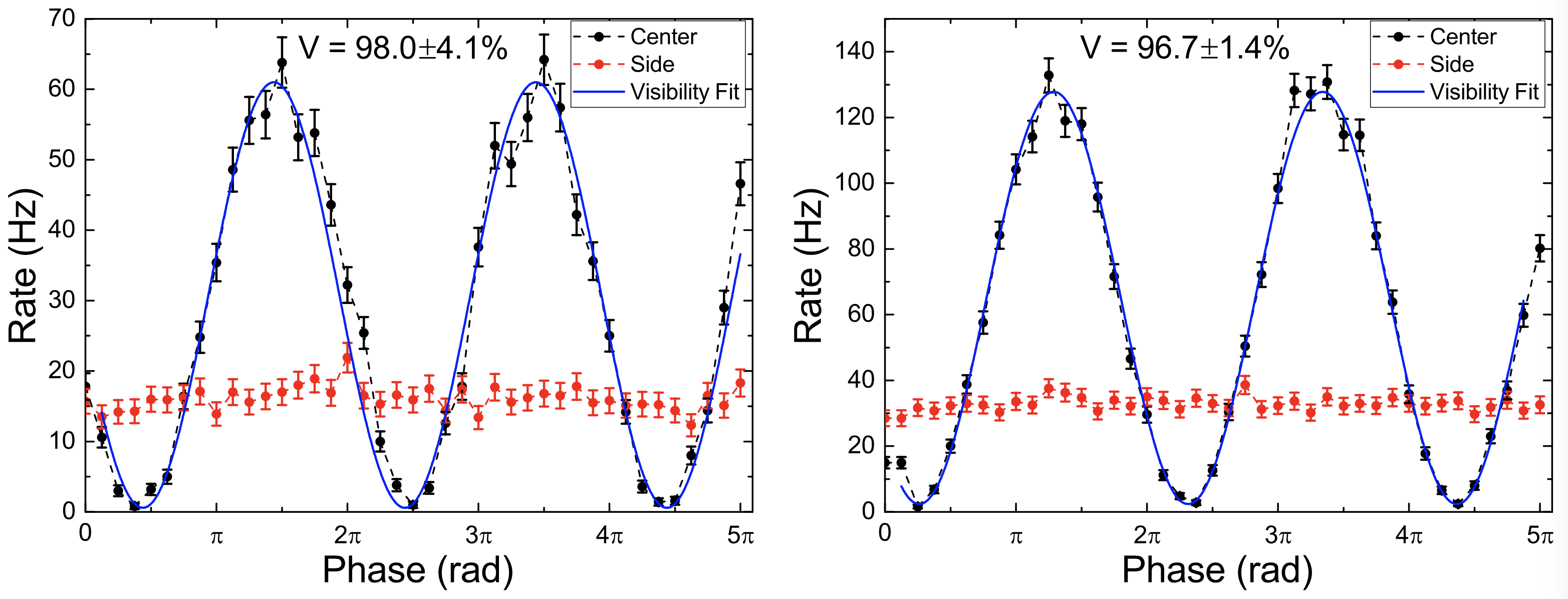}
%	\centering
%	\begin{minipage}{.5\textwidth}
%	  \centering
%	  \includegraphics[width=\linewidth]{FigRateTest5v2.eps}
%	\end{minipage}%
%	\begin{minipage}{.5\textwidth}
%	  \centering
%	  \includegraphics[width=\linewidth]{FigRateTest9v2.eps}
%	\end{minipage}
	\caption[width=\linewidth]{Plots of the coincidence rates (blue curve) for the central (Region (1) in Fig. \ref{fig:coins}b) and the average of the side peaks (Regions (2) and (3) in Fig. \ref{fig:coins}b). Each point is obtained with a 5 seconds integration time and the step increment of $\pi/8$. The side peak rates show the stability of the measurement. Here noise counts are not removed from the measurements, (a) corresponds to the 2-FSR shift resonances with a noise of $0.5$ counts/s, (b) is the 3-FSR shift resonances with a noise of $0.9$ counts/s. The error bars for all points come from Poissonian statistic.} 
	\label{fig:vis}
\end{figure*}

The photon pairs are genuinely energy-time entangled as they are produced by SFWM \cite{azzini2012classical,jiang2015silicon,grassani2016energy,wakabayashi2015time,preble2015chip,silverstone2015qubit,silverstone_-chip_2013,faruque2018chip}. Entanglement is analyzed using a folded Franson arrangement consisting of an unbalanced fiber Michelson interferometer (F-MI) (Fig. \ref {fig:schema}d) \cite{franson1989bell,thew2002experimental}. A piezo-transducer is used to extend the fiber in one arm, changing the imbalance of the interferometer, and thus the relative phase between the two arms. Energy-time quantum correlation are revealed by the coherent superposition of the contributions coming from identical two-photon paths (short-short and long-long) contrarily to the contributions coming from different paths (long-short, or conversely). Consequently, a coincidence histogram with the emergence of 3 peaks is recorded~\cite{aktas}(Fig. \ref{fig:coins}b). The central peak gathers the two indistinguishable contributions leading to interference, provided all experimental conditions reported in the Methods section are satisfied. The total and average numbers of coincidences in the central and side peaks, respectively, (Fig. \ref{fig:coins}(b)) are used to plot the patterns shown in Fig. \ref{fig:vis}.

\begin{figure*}
	\centering
	\includegraphics[scale=0.7]{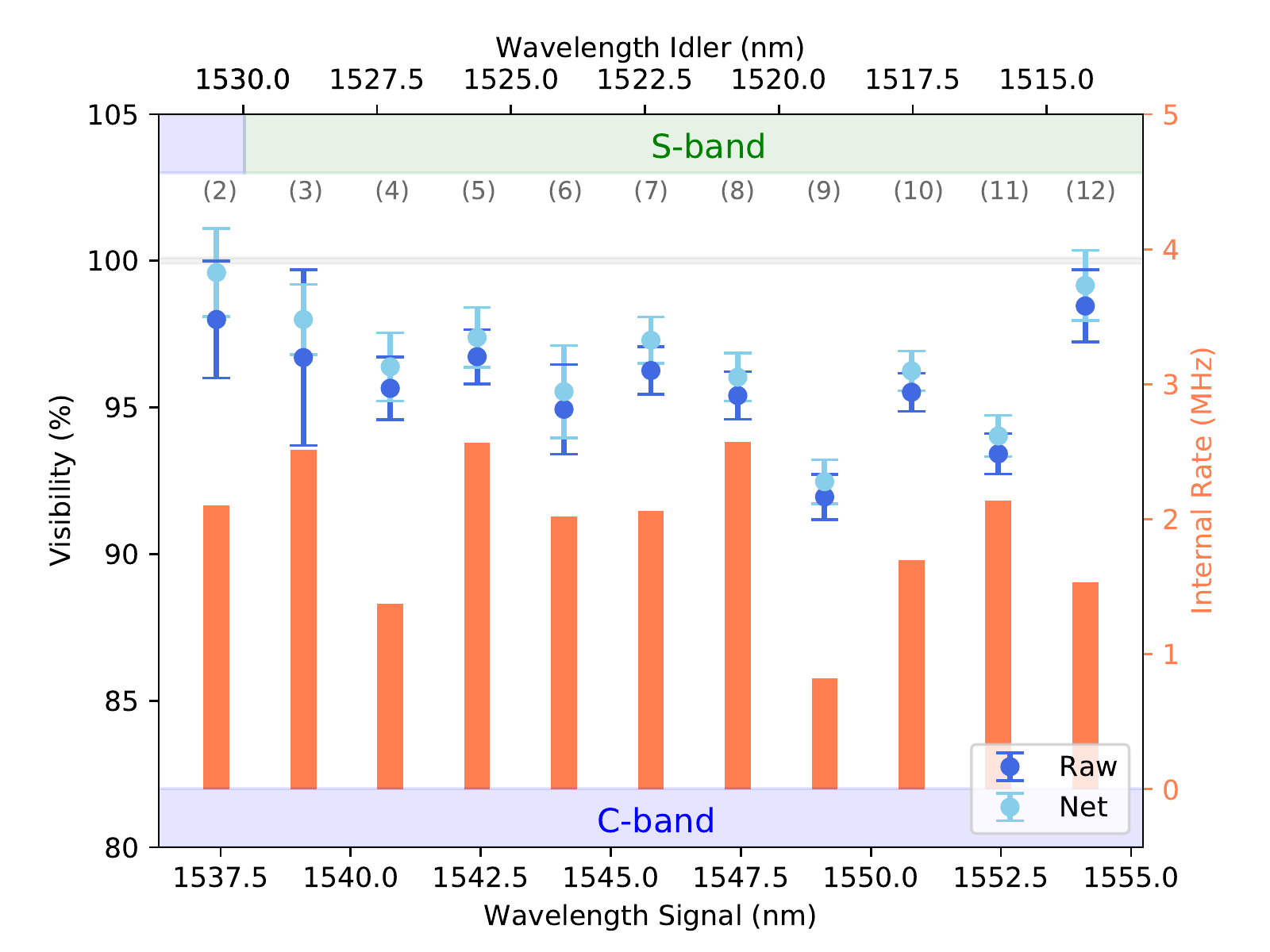}
	\caption{Raw and net visibilities and internal rate plotted as a function of the signal and idler wavelengths. For sake of clarity, we associate signal and idler wavelengths to their corresponding telecom band.}
	\label{fig:allvisibility} 
\end{figure*}

Interference patterns recorded for 2-FSR shift (Fig. \ref{fig:vis}a) and 3-FSR shift (Fig. \ref{fig:vis}b) resonances are obtained with a phase resolution of $\frac{2\pi} {20}$, compliant with the F-MI stability. The interferometer $\pi$-dephasing time-scale is in the hour range which leaves enough time to perform a scan without being subjected to detrimental phase drifts. The side peak in Fig \ref{fig:coins}(b) also shows the stability of the photon generation process. The typical acquisition time for recording two fringes is 6 minutes. The noise figure in the central peak is of about $0.5$ and $0.9$ counts/s for the 2-FSR shift and 3-FSR shift resonances, respectively.
The two-photon interference fringes are fitted with respect to $N_0(1- Vcos(2\phi))$, where $N_0$ is the mean number of coincidences and $V$ the fringes visibility considered as a free-fit parameter to infer the visibility. Raw visibilities of (98.0 $\pm 2$)\% and (96.7$\pm 3$)\% without subtraction of photonic or detector noise are obtained, for the 2-FSR shift pairs (R-squared of 0.96) and for the 3-FSR shift pairs (R-squared of 0.99), respectively. The net visibility is obtained by subtracting photonic noise originating from the detectors’dark counts (200 counts/s) and corresponds to (99.6 $\pm 1.5$)\% and (98.0 $\pm 1.2$)\% for the 2-FSR shift and 3-FSR shift pairs, respectively.\\
We extend our investigations according the same methodology for subsequent paired-channels within the S-band and the full C-band~\cite{telecom}. More precisely, we explore the entanglement quality of 9 extra paired-channels, \textit{i.e} up to 11-FSR shift, spaced by 40 nm (1515-1555~nm) on both sides of the pump channel, leading to the ability of supporting a high number of users in a multiplexing scenario\cite{aktas}. The details for the signal and idler wavelengths corresponding to i-FSR shift, with $2 \leq i \leq 11$, are reported in the Methods. The raw visibilities for all the paired-channels exceeds 92\% for an internal rate $\ge$ 1~MHz as shown in Fig.~\ref{fig:allvisibility}.\\
Note two other photonic noise contributions have been evaluated before being neglected: multiple photon-pair events and non-perfect overlap between the two identical two-photon paths ("short-short" and "long-long"). The former was not considered because of the low mean number of photon pairs per gate window ($\bar{n}=3.10^{-4}$). The origin of the latter contribution lies in the wavelength difference between the signal and the idler over the full range of analysis ($\sim$ 50~nm), leading to potentially slightly distinguishable "short-short" and "long-long" two-photon paths. The time arrival difference has been evaluated to be $<10^{-5}$ ps whereas the full width at half maximum of the coincidence peak is equal to 160~ps (Fig. \ref{fig:coins}a), which represents several orders magnitudes higher than the shift between the "short-short" and "long-long" path.\\

This result not only stands as among the highest raw quantum interference visibility for time-energy entangled photons from a micro/nanoscale integrated circuit over such a large spectral window (partially over the S band and fully over the C band) but also the first entanglement qualification of a complex integrated circuit including the pump filter~\cite{Wakabayashi:15,grassani_micrometer-scale_2015, rogers_high_2016}. Furthermore, generating genuinely a pure maximally entangled state from an integrated source associated with a high coincidence rates will be of special interest for QIS experiments.

\section*{Discussion}

Our strategy is geared towards developping "Plug-\&-Play" and scalable quantum photonic systems by exploiting sophisticated silicon-based architectures combined with off-the-shelves telecom components \cite{popovic_multistage_2006,luo_silicon_2012,dong_ghz-bandwidth_2010, piekarek_high-extinction_2017, harris_integrated_2014}. 
Notably, the high SNR (>60) and the low noise observed in both coincidence and interference experiments (Fig. \ref{fig:coins} and \ref{fig:vis}) show the effectiveness of our pump rejection solution, achieved in an all-passive manner. Furthermore, the acquisition time for recording the interference patterns is on the order of several minutes, allowing a setup free of complex stabilization systems for the F-MI. Note that the on-chip pump rejection is complemented by two bandpass filters, corresponding to a total amount of 82\;$\mathrm{dB}$, which is lower than the typical necessary value of 100\;$\mathrm{dB}$~\cite{piekarek_high-extinction_2017}. This difference probably arises from an underestimation of the real rejection of the on-chip pump filter, which is closer to 80\;dB. This assumption is confirmed by a previous work based on the design, the implementation and the qualification of such a filter\cite{oser2018coherency}. Moreover, signal and idler photons are coupled to off-chip bandpass filters leading to spatial filtering of the light scattered from the chip. 

We also stress that, the modal coupler cascaded with the PF allows the exploitation of the rejected signal, in order to match the frequency combs from the RR with the pump laser (Fig. \ref{fig:schema}a). We could use this strategy to demultiplex signal and idler photons in an all-passive way by cascading several modal couplers of this type. In this case, the period of the gratings that are part of the PF has to be adjusted according to the signal and the idler wavelengths. Furthermore, the slope of the edges corresponding to the transition from high blocking to high transmission of the PF (Fig.~\ref{fig:spect}) is evaluated to be of 22.3~dB/100~pm, being therefore compatible with narrower spectral channels. Hence, the exploitation of a micro-ring associated with a 50~GHz FSR instead of 200~GHz stands as a relevant improvement in order to enhance the number of users supported by the chip up to 44~pairwise users~\cite{imany_50-ghz-spaced_2018}. This solution would neither impact the design of the PF nor the demultiplexing stages as 50~GHz stands as a standard for the ITU channel grid.% Finally, our entanglement qualification over channel pairs could be extended to other telecom bands provided appropriate DWDM components. We believe that this approach paves the way of a global quantum network without any active switching \cite{ursin}.

In summary, we have demonstrated near perfect entanglement quality out of a single photonic chip embedding both generation and pump rejection building blocks. We have measured raw visibilities exceeding 92~\% for 11 channel pairs over telecom bands covering the spectral range 1515-1555~nm. Our approach, combining high performance, flexibility, scalability and compliance with telecom standards, stands promising for operational QIS applications. This brings an essential step closer to demonstrate ambitious photonic quantum devices enabling on-chip generation, filtering and manipulation of quantum states.

\section*{Methods}

\paragraph*{Design}
 
 The ring resonator is in a racetrack configuration with a radius of $49 \; \mathrm{\mu m}$ and a straight section of $18 \;\mathrm{\mu m}$. The gap between the ring and the bus waveguide is $150 \;\mathrm{nm}$ wide. The ring  waveguide width is $600 \;\mathrm{nm}$, making it multi-mode, whereas the bus waveguide width is $400 \;\mathrm{nm}$, to be single-mode. 
%With a  With only minor adjustments such ring could be aligned with the ITU frequency grid. %

The pump filter is composed of 9 cascaded sections of multi-mode Bragg filters (BF), each being $300 \;\mathrm{\mu m}$ long. The corrugation of the filters is $W_{corr} = 225 \;\mathrm{nm}$, the waveguide width and the pitch of each section are equal to $W = 1.15 \;\mathrm{\mu m}$, and $P = 290 \;\mathrm{nm}$, respectively, with a duty cycle of 50 \%. Tapers and apodizing of the corrugation have been placed at the ends of all sections to improve transmission. Consequently, the losses of our filter comes mainly due to the propagation and are as low as 2.5dB. Furthermore, our device works fully passively.
Finally, the grating couplers are optimized to reduce Fabry-Perot ripples in the collected signal for an easier analysis of the transmission spectrum.

\paragraph*{Fabrication}
The circuit was made using silicon-on-insulator wafers comprising a $220 \;\mathrm{nm}$ thick silicon and a $3 \;\mathrm{\mu m}$ buried oxide layer. Electron beam lithography (Nanobeam NB-4 system $80kV$, with a step size of $5 \;\mathrm{nm}$), dry, and inductively coupled plasma etching (SF6 gas) were used to define the patterns.
Note that all fabrication steps, notably concerning the dimensions of the components (minimum feature size of $145 \;\mathrm{nm}$ from the pitch), can be realized with standard deep-UV lithography. 
%Also a metallic heater, deposited on top of the resonator, would have permitted fine alignment of the generated frequency response of the ring and filter.

\paragraph*{Measurements}
The full experimental setup is presented in Fig. \ref{fig:schema}. Light from a narrowband tunable telecom CW laser (Yenista Tunics Plus laser) is coupled to the device. Before the silicon chip, a tunable bandpass filter (Yenista XTM-50) is used to clean the laser from the amplified spontaneous emission (ASE), which would otherwide be coupled to the chip and not filtered by any other components. Then a polarization controller is used to set the laser light to the TE polarization mode (Fig. \ref{fig:schema}a). The TM polarization is suppressed by both grating couplers (about $40 \;\mathrm{dB}$ each) of the circuit as well as by the waveguide itself due to the asymmetric cladding (silicon waveguide between silica and air layers). Finally, a Peltier driven by a temperature controller (Thorlabs TED 200C) is used to ensure a thermal regulation of the sample to $(21 \pm 0.01)\;\mathrm{^{\circ}C}$.

For the time correlation measurements, we employ, after the sample, a beam splitter followed in each arm by a bandpass tunable filter (OZ optics) of $600 \;\mathrm{pm}$ bandwidth and $20 \;\mathrm{dB}$ extinction ratio (Fig. \ref{fig:schema}a). This is for demultiplexing signal and idler photons, one filter being set to the signal wavelength and the other to the idler wavelength (Fig. \ref{fig:schema}e). %It is a very inefficient demultiplexing as in addition to the losses, only a fourth of the pair can be counted.
We use two superconducting single photon detectors (SSPD, ID Quantique ID281) connected to a TDC (PicoQuant HydraHarp 400) for recording coincidence counts with a bin precision of $1 \;\mathrm{ps}$. 

Exploiting energy-time observable relies on the systematic lack of information of the pairs' creation times within the coherence time of the employed CW pump laser. In practice, the twin photons pass through the unbalanced interferometer (see Fig. \ref{fig:schema}d) following either the same path (short-short or long- long) or different paths (long-short, or conversely) \cite{aktas}. These contributions are distinguished by measuring the arrival times of the idler photons with respect to those of the signal photons using the TDC. This enables recording a coincidence histogram comprising three peaks (Fig. \ref {fig:coins}b). The side peaks (labeled (2) and (3)) correspond to the situations where the paths are different whereas the central peak (labeded (1)) gathers the two indistinguishable cases (short-short and long-long) leading to two-photon interference. Note that the F-MI imbalance ($\sim$ 350\;ps) needs to be greater than the coherence time of the single photons ($\sim$ 100\;ps) to avoid first-order interference and shorter than the coherence time of the CW pump laser ($\sim$ 100\;ns) in order to have a coherent superposition between short-short and long-long contributions in the central peak. By using a narrow coincidence window that excludes the side peaks, entanglement can be revealed according to the coincidence counting evolution $N_0(1- Vcos(2\phi))$, where V and $\phi$ represent the fringe visibility and the phase accumulated by the single photons in the interferometer, respectively. The visibility stands as our the main figure of merit, and correlation described by such a coincidence function with a visibility higher than 70.7 \% cannot be described by any local hidden-variables theories \cite{Tanzilli2002}.
%\paragraph*{Energy-time entanglement and visibility}

\begin{table*}[htbp]
  \centering
  \caption{\bf Summary of the losses experienced by the paired photons}
  \resizebox{.7\textwidth}{!}{%
  \begin{tabular}{cccc}
  \hline
  Components & Signal (dB) & Idler (dB) & Total (dB) \\
  \hline
  Anti-ASE filter & & & $7 \pm 0.1$ \\
  Polarizer controller & & & $0.3 \pm 0.1$ \\
  Grating coupler & & & $5 \pm 0.2$ \\
  \hline
  Integrated filter & $2 \pm 0.1$ & $2 \pm 0.1$ & $4 \pm 0.2$ \\
  Grating coupler & $5 \pm 0.2$ & $5 \pm 0.2$ & $10 \pm 0.4$ \\
  Beam splitter & $6 \pm 0.2$ & $6 \pm 0.2$ & $6 \pm 0.2$ \\
  Bandpass filters & $2 \pm 0.5$ & $2.5 \pm 1.5$ & $4.5 \pm 2$  \\
  SSPD & $2 \pm 0.2$ ($\sim 60\%$) & $3 \pm 0.2$ ($\sim 50\%$)& $5 \pm 0.4$ \\
  \hline
  Total & $17 \pm 1.2$ & $18.5 \pm 2.2$ & $35.5 \pm 3.2$ \\
  \hline
  \end{tabular}}
  \label{tab:loss}
\end{table*}

\begin{table*}[htbp]
  \centering
  \caption{\bf Summary of signal and idler wavelengths. The pump is set to $\lambda_p$~=~1534.2~nm.}
  \begin{tabular}{c|ccccccccccc}
  \hline
  FSR & 2 & 3 & 4 & 5 & 6 & 7 & 8 & 9 & 10 & 11 & 12 \\
  \hline
  $\lambda_{idler}$ (nm) & 1530.8 & 1529.1 & 1527.4 & 1525.8 & 1524.2 & 1522.5 & 1520.9 & 1519.2 & 1517.6 & 1516.0 &  1514.4 \\
  \hline
  $\lambda_{signal}$ (nm) & 1537.4 & 1539.0 & 1540.8 & 1542.4 & 1544.1 & 1545.8 & 1547.5 & 1549.2 & 1550.9 & 1552.6 & 1554.3 \\
  \hline
  $V_{raw}$ (\%) & 98.0 & 96.7 & 95.6 & 96.7 & 94.9 & 96.3 & 95.4 & 91.9 & 95.5 & 93.4 & 98.5 \\
  \hline
  $V_{net}$ (\%) & 99.6 & 98.0 & 96.4 & 97.4 & 95.5 & 97.3 & 96.0 & 92.5 & 96.2 & 94.0 & 99.2 \\
  \hline
  Rate (MHz) & 2.1 & 2.5 & 1.4 & 2.6 & 2.0 & 2.0 & 2.6 & 0.8 & 1.7 & 2.1 & 1.5 \\
  \hline
  \end{tabular}
  \label{tab:loss}
\end{table*}

\section*{Acknowledgment}
The authors thank Bhaskar Kanseriin for fruitfull discussion and valuable inputs, especially for the last measurements.\\
This work has been conducted within the framework of the project OPTIMAL granted by the European Union by means of the Fond Européen de développement regional (FEDER). The authors also acknowledge financial support from the Agence Nationale de la Recherche (ANR) through the projects SITQOM (grant agreement ANR-15-CE24-0005) and the French government through its program “Investments for the Future” under the Université Côte d’Azur UCA-JEDI project (under the label Quantum@UCA and OPENING) managed by the ANR (grant agreement ANR-15-IDEX-01).

The sample fabrication has been performed at the Plateforme de Micro-Nano-Technologie/C2N, which is partially funded by the "Conseil G\'en\'eral de l'Essonne". This work as partly supported by the French RENATECH network. The authors also acknowledge technical support from IDQ.

\section*{Author contributions}

F.M. and D.O. performed the experiments. C.A.R. and D.O. designed the circuit. X.L. and D. O. fabricated the sample. S.T. and L.L. designed the experiment. G.S., X.H. and O.A. performed calibrations and technical support for the detectors. S.T., L.L., and D.O. wrote the paper with inputs from L.V. and E.C.

%%%%%%%%%%%%%%%%%%%%%%% References %%%%%%%%%%%%%%%%%%%%%%%%%

%%%%%%%%%% If using BibTeX:
%\bibliography{biblio}


\end{document}